%Paper: hep-ph/9204231
%From: pvl@surya11.cern.ch (Peter Landshoff)
%Date: Thu, 23 Apr 92 10:05:34 +0200

------------------------------------------------------------------------------
%figures are appended in a separate file

\magnification\magstephalf

\font\rfont=cmr8
\def\ref#1{$^{\hbox{\rfont {[#1]}}}$}

  %%Fonts

\font\fourteenbf=cmbx12 scaled\magstep1

\font\twelverm=cmr12
\font\twelvebf=cmbx12

  %%Greek
   
  \def\o{\omega}

\def\pd {\partial}
\def\pmb#1{\setbox0=\hbox{#1}% \kern-.025em\copy0\kern-\wd0
 \kern.05em\copy0\kern-\wd0 \kern-.025em\raise.0433em\box0 }

  %%Fractions
\def \half {{\scriptstyle {1 \over 2}}}

\def \quarter {{\scriptstyle {1 \over 4}}}

\def \sixth {{\scriptstyle {1 \over 6}}}
 %

 %%FORMATTING

\def \cl {\centerline}
\parskip=6pt
\parindent=0pt
\hsize=17truecm\hoffset=-5truemm
\voffset=-1truecm\vsize=26.5truecm
\def\footnoterule{\kern-3pt
\hrule width 17truecm \kern 2.6pt}

{\nopagenumbers

\rightline{CERN-TH-6384/92}
\vskip 2truein
\cl{{\fourteenbf REACTION RATE IN A HEAT BATH}}
\bigskip
\bigskip
\cl{{\twelverm M Jacob and P V Landshof{}f}\footnote{$^{\dag}$}{On leave of
 absence from
DAMTP, University of Cambridge}}
\medskip
\cl{CERN, Geneva}
\vskip 1truein
{\twelvebf Abstract}
\bigskip
{\twelverm
We show in detail how the presence of a heat bath of photons effectively gives
charged particles in the final state of a decay process a
temperature-dependent mass,
and changes the effective strength
of the force responsible for the decay. At low temperature,
gauge invariance causes both these
effects to be largely cancelled by absorption of photons from the heat bath
and by stimulated emission into it, but at high temperature the
temperature-dependent mass is the dominant feature.
}
\vskip 3truein
CERN-TH-6384/92

February 1992
\vfill\eject
\vfill\eject
}
\pageno=1
\bigskip
\bigskip
{\bf 1 Introduction}

In thermal field theory, one calculates thermal averages of observables
$Q$:
$$
\langle Q \rangle = Z^{-1} \sum \langle i| e^{-\beta H}Q|i\rangle
\eqno(1)
$$
where the sum extends over a complete set of physical states.
As usual, $H,Z$ and $\beta$ are respectively the Hamiltonian, the grand
partition function and the inverse temperature of the system.
The perturbation theory that can be developed to calculate such quantitities
$\langle Q \rangle$
is very similar to the familiar Feynman perturbation theory\ref{1}. Thus it
seems natural  to suppose that  concepts from Feynman perturbation theory
apply also to calculations in the presence of a heat bath.  A particular
example is the idea of mass renormalisation and
the supposition that thermal effects somehow give a particle an additional
``thermal'' mass, though there has been disagreement about how exactly
this additional mass should be defined\ref{2}. In this paper, we derive
the mass correction, and consider the other
effects that the heat bath has on reaction rates.

Since our aim is to elucidate basic principles and methods, we study a very
simple problem.
We consider a heat bath of photons in equilibrium at temperature $T$
and a neutral scalar particle $\Phi$ that decays\footnote{$^{\dag}$}
{Our results apply equally
well when the decaying particle has spin. As is
customary, we shall ignore the problem that strictly-speaking there is
no proper field theory of unstable particles. We could equally have used
as our example the scattering process $\Phi\Phi\to\bar\phi\phi$ through
a  weak interaction $g\Phi ^2\phi ^{\dag}\phi$.}
into a charged scalar
particle and its antiparticle. We take the decay interaction to be
through the weak pointlike coupling $\lambda\Phi\phi ^{\dag}\phi$
and work to lowest
order in $\lambda$. For simplicity, initially we do not work with proper
scalar electrodynamics, but pretend that the photons are scalar particles
with a mass $\mu$ and the simple interaction
$$
(eA+e^2A^2)\phi ^{\dag} \phi
$$
Later, we go over to the proper gauge theory.

We suppose that the heat bath has not been in existence for a time long
enough for
the particles  $\Phi$ and $\phi$ to be anywhere near  thermal
equilibrium; rather, the heat bath will be supposed to consist only of the
photons. This is again because we want to consider the simplest possible
problem in order to concentrate on basic principles. As we describe briefly
at the end of section 3,  it is straightforward
to extend the analysis so as to include the particles $\Phi$ and $\phi$
in thermal equilibrium in the heat bath.

In section 2
we show in detail how the heat bath effectively gives the charged particles
an additional squared mass of order $e^2T^2$, and how it also changes the
 effective
value of the decay coupling constant $\lambda$. Further, there are effects
arising from absorption of photons from the heat bath, and stimulated
emission of photons into it.

When we go over to the proper gauge theory in section 3,
additional features enter.
This is that gauge invariance ensures that there are various cancellations.
Part of the cancellation is of the infrared divergences that are now
present in the change in the effective value of $\lambda$ and in the
absorption and stimulated emission\ref{5}. But we find that it is not only the
divergences that cancel: in the low-temperature expansion up to
order $e^2T^2$ everything cancels, and the
net effect of the heat bath on the decay rate is zero up to this order.
However, at high temperature the mass-shift effect is of order $e^2T^2$,
while the other effects together are only of order $e^2T$, so that the mass
shift is the most important feature.
\bigskip
{\bf 2 Change of effective mass and coupling}

We calculate to order $e^2$ the difference between the decay rate in
the heat bath and that at zero temperature (which is the same as that with no
heat bath, as calculated in ordinary Feynman theory). It is convenient to
work with real-time thermal perturbation theory for the electromagnetic
interaction. Normally, in real-time
thermal perturbation theory there are complications associated with field
doubling\ref{3}, but these are not encountered when the heat bath contains
only photons.

The derivation of the real-time perturbation theory requires an adiabatic
switching off of the interaction with the heat bath at asymptotic
times\ref{4}. We impose the switching off in the rest frame of the heat bath
(though the choice of frame for this probably is not essential). We do not
assume that the decaying particle is necessarily at rest in this frame.
We shall find it necessary to retain explicitly the
switching factor in the first part of our calculation. We shall take it
to be $S_{\eta}(t)$. We also need to introduce a switching factor $S_{\xi}(t)$
for the interaction that causes the decay of the $\Phi$ particle.
Our results will not be sensitive to the precise
shape of these functions, provided they are  very close to unity for all finite
$t$ and go to zero when $|t|\to \infty$. It will be helpful to have in mind
the explicit forms $S_{\eta}(t)=e^{-\eta |t|}$ and $S_{\xi}(t)=e^{-\xi |t|}$,
with Fourier transforms
$$
S_{\eta}(\omega )=\int {{dt}}\; e^{i \omega t} e^{-\eta |t|} =
   {{2\eta}\over{\omega ^2 +\eta ^2}}
$$ $$
S_{\xi}(\omega )=\int {{dt}}\; e^{i \omega t} e^{-\xi |t|} =
   {{2\xi}\over{\omega ^2 +\xi ^2}}
\eqno(2)
$$
Eventually, we let $\eta ,\xi \to 0$, and then both functions in (2)
become $2\pi\delta (\omega )$.
But before we take that limit, at each vertex of our Feynman graphs
there will be an additional incoming energy $\omega$, with a corresponding
multiplicative factor (2) and an integration over $\omega$.
The heat bath is coupled to the decay process for a very long time when
$\eta$ is very small, but as it is nevertheless switched off in the
remote past and future the mass of the $\phi$ particle in the asymptotic
{\it in} and {\it out} states is just
its zero-temperature
value $m$.  We show that  nevertheless the self-energy insertions
generated by the heat bath effectively change this value in the formula
for the decay rate, and derive the  other effects caused by the heat bath.

A change of mass is, of course, usually associated with renormalisation.
We recall that the familiar
zero-temperature renormalisation is partly a mere  rearrangement of
the unrenormalised perturbation expansion, but also it involves making a
change in it: one introduces a new choice of
the asymptotic condition to be satisfied by the fields. The mere rearrangement
concerns how one deals with the self-energy insertions on internal-line
propagators in an amplitude, and has two aspects.  A part $\delta m^2$
of the self-energy $\Pi(p^2)$ is
subtracted from $\Pi$ and added to the squared mass:
$m^2 \to m_R^2=m^2+\delta m^2$; this subtraction is
chosen such that the new $\Pi$ vanishes at
$p^2=m^2_R$, so that the propagator pole is at this point.
Also, $Z=(1-\Pi '(m^2_R))^{-1}$ is factored off
from the propagator and the square root
of this factor is instead  made to multiply the coupling constant at
each end of the propagator. If all the vertices of the Feynman graph were
internal vertices, this mass and coupling-constant renormalisation would
be just a rearrangement that does not change the Feynman integral. But
in order to justify the same change of coupling
constant at external vertices, where there are a smaller number of
propagator lines attached than at internal vertices, the asymptotic
condition must be changed
so that the normalisation  of the $in$ and $out$ free fields
is changed by a factor $\surd Z$. Also, these fields are required to
have mass $m_R$  instead of the original mass.
This change of
asymptotic condition, being more than just a  rearrangement, must be
given physical justification; in electrodynamics one usually does this
by noting that the lowest-order electron-electron scattering
amplitude at small momentum transfer, renormalised in this way,
agrees with the Rutherford formula, say,
when one identifies the renormalised mass and charge with those
listed in the data tables. Happily, the renormalisation absorbs the infinities
of the theory into the renormalised quantities, but one would need to go
through
such a procedure even if there were no infinities. There are many other
renormalisation procedures that absorb the infinities equally well, but
the one we have described is the one that most directly makes contact
with the data-table values of the physical quantities.

There is obviously no reason why one should not make an exactly similar
{\it rearrangement} in thermal perturbation theory, where now the self-energy
has an additional temperature-dependent part, since it does not change
anything. That is, one is free to rearrange the calculation such that
a new temperature-dependent mass and a new temperature-dependent coupling
are used on internal lines and vertices. This mass will correspond to
the position of the pole in external legs of the Green's functions of
the finite-temperature theory, and the coupling will correspond to a pole
residue. But some justification is needed
if one wants to assume that these new quantities enter in a simple way
in the calculation of
finite-temperature reaction rates.
We shall show in this paper that in fact one does not need to make any
new assumptions to justify this.

We set up the perturbation theory in the interaction picture that
coincides with the Heisenberg picture in the remote past; that is,
we work with {\it in} fields and states. Although the heat bath is
not interacting with the particle $\Phi$ initially, it is nevertheless
present and so we should describe the initial system by the density matrix
$$
\rho=Z_0^{-1}\sum _i e^{-\beta H_0^{\gamma}}
 a_{\Phi} ^{\dag}(P)\;|i\rangle \langle i|\;a_{\Phi} (P)
\eqno(3)
$$
where $|i\rangle$ are the possible states of the heat bath of noninteracting
photons and $H_0^{\gamma}$ and $Z_0$ are its Hamiltonian and
grand partition function.
The decay probability is
$$
\sum _f \langle f| T\rho T^{\dag}|f\rangle =
Z_0^{-1}{\hbox{tr }} e^{-\beta H_0^{\gamma}}
a_{\Phi} (P)T T^{\dag} a_{\Phi} ^{\dag}(P)
\eqno(4)
$$
where we have used the completeness relation for the states $|f\rangle$ and
the trace is in the subspace of heat-bath states.
We insert the standard perturbation-theory operator expression for the
transition matrix $T$, using the {\it in} fields,
and evaluate the terms of (4) that are of order
$e^2 \lambda ^2$. This results in the types of diagrams of figure 1.
In these diagrams the charged-particle lines correspond to the usual
zero-temperature propagators and $\delta$-functions, with the
zero-temperature data-table masses,  because the heat
bath is supposed not to contain any charged particles.
The photon lines correspond to
$$
\Delta (k;\beta )=\int d^4x\; e^{-ik.x}Z_0^{-1}{\hbox{tr }}
  e^{-\beta H_0^{\gamma}}A(x)A(0)
\eqno(5a)
$$
for the real-photon lines and
$$
D(k;\beta )=\int d^4x\; e^{-ik.x}Z_0^{-1}{\hbox{tr }}
  e^{-\beta H_0^{\gamma}}{\hbox{T}}\left (A(x)A(0)\right )
\eqno(5b)
$$
for the virtual ones (with T the real-time time-ordering operator).
(Recall that, initially, we are taking the ``photons'' to be scalar particles
with mass $\mu$.)

Because we want to calculate the difference between the finite-temperature
and zero-temperature decay rates, we subtract from (5) the zero-temperature
expressions, namely
$$
\Delta  (k;\infty )=
  \theta (k_0) 2 \pi  \delta (k^2-\mu ^2)
\eqno(6a)
$$
and
$$
D(k;\infty )=
{{-i}\over {k^2-\mu ^2+i\epsilon}}
\eqno(6b)
$$
After this subtraction, we are left with
$$
\Delta ^{\beta}  (k)=D ^{\beta}(k)=
  2 \pi \delta (k^2-\mu ^2)f^{\beta}(k_0)
\eqno(7)
$$
with
$$
f^{\beta}(k_0)={1\over{e^{\beta |k_0|}-1}}
\eqno(8)
$$
Although these two functions are equal, they represent different physical
processes: $\Delta ^{\beta}$ corresponds to permanent absorption or emission
of a photon from or into the heat bath, while $D ^{\beta}$ represents
absorption together with emission (or vice versa).
Notice that, unlike at zero temperature, the real-photon function
$\Delta ^{\beta} $ does not contain a factor $\theta (k_0)$; this is
because it describes both emission of real photons into the heat bath
and absorption from it, depending on the sign of $k_0$.

The additions to the two self-energies appearing in figures 1a
and 1b arising from the heat bath are
$$
\Pi _a^{\beta}(p_1)=e^2\int {{d^4k}\over{(2\pi )^4}}D ^{\beta}(k)$$
$$
\Pi _b^{\beta}(p_1)=e^2\int {{d^4k}\over{(2\pi )^4}}D ^{\beta}(k)
{1\over{(p_1+k)^2-m^2+i\epsilon}}
\eqno(9)$$
(Actually, $\Pi _a^{\beta}$ is independent of $p_1$.)

At each vertex of the graphs of figure 1 there is conservation of
3-momentum. However, we must allow for a spurion energy $\o$, associated
with the switching factors (2).
Although eventually $\eta ,\xi  \to 0$ so that
the spurions then disappear, we need them initially to make sense of
the first two graphs. Without the spurion energies, the
internal (zero-temperature) scalar-particle propagator would be
$(p^2-m^2)^{-1}$
which then has to be evaluated at its pole because there is also an
on-shell line with the same momentum. But before $\eta ,\xi \to 0$ the
on-shell line's momentum is not the same as that of the internal-line
propagator
and so the pole contribution
is regularised.
It will be a useful check on our final answer that
it is independent of the relative rates at which $\eta ,\xi \to 0$, or
indeed of the precise shapes of the switching-off functions.

Each graph of figure 1 represents an amplitude on the left-hand side
of the cut lines and the complex conjugate of an amplitude on the right.
Denote the spurion energy leaving the left-hand decay vertex by $\Omega$
and that entering the right-hand one by $\Omega '$. Similarly, introduce
energies $\omega$ for the spurion energies that leave the heat-bath-interaction
vertices to the left of the cut lines, and energies $\omega '$ for those that
enter those to the right. The squared matrix element contains, among other
things,
$$
\int {{d\Omega}\over{2\pi}} {{d\Omega '}\over{2\pi}}
\prod\left ({{d\omega}\over{2\pi}}\right )
\prod\left ({{d\omega '}\over{2\pi}}\right )
2\pi\delta \left (\Omega +\sum (\o )-\Delta E\right )
2\pi\delta\left  (\Omega ' +\sum (\o ' )-\Delta E\right  )
\phantom{xxxxxxxxxx}
$$
$$
\phantom{xxxxxxxxxx}
S_{\xi}(\Omega )S_{\xi}^*(\Omega ')
\prod S_{\eta}(\omega )\prod S^*_{\eta}(\omega ')
\eqno(10)
$$
where the switching-off functions $S$ are defined in (2) and
$\Delta E$ is the difference between the initial energy and the
sum of the energies of the cut lines.

When we calculate a probability from a squared $T$-matrix element,
we normally have a factor\hfill\break
 $(2\pi )^4\delta ^{(4)}(\Delta P)$
in both the amplitude and the complex-conjugate amplitude, where
$\Delta P$ is the difference between the total 4-momentum on the cut lines
and that on the external lines.
In order to obtain a cross-section we normally factor off
$(2\pi )^4\delta ^{(4)}(0)$, which is interpreted as the volume
times the total interaction time. This last has been affected
by our introduction of the switching factors: while we still have
 $(2\pi )^3\delta ^{(3)}(\Delta {\bf P})$
in both the amplitude and the complex-conjugate amplitude,  the
square of $2\pi \delta (\Delta E)$ is no longer there.
In order to see what it
has become, consider a graph that does not ``need'' the switching factors,
that is one for which we may set the $\omega$ and $\omega '$ to zero
in the propagator factors before performing the integrations in (10).
The integral (10) is then equal to
$$
\int dt dt' e^{i\Delta E\;(t-t')}
 S_{\xi}(t)S_{\xi}(t')\prod S_{\eta}(t)\prod S_{\eta}(t')
\eqno(11)
$$
In the limit where the switching factors $S_{\eta}$ and $S_{\xi}$
become equal to unity this is just $2\pi\delta (\Delta E)2\pi\delta (0)$.
Before we take this limit, (11) is small except very near $\Delta E=0$,
that is it is almost equal to  $2\pi\delta (\Delta E)$ times a large factor $N$
obtained by integrating (11) with respect to $\Delta E$:
$$
N_n=\int dt [S_{\xi}(t)]^2 [S_{\eta}(t)]^n
\eqno(12a)
$$
where $n$ is the total number of heat-bath vertices.
It is this integral that we must identify as the total
interaction time and factor it off. In the case of the
simple explicit forms (2) for the
switching functions it is
$$ N_n={2\over{2\xi +n\eta}}
\eqno(12b)
$$
where  we have omitted
terms that remain finite or vanish when $\xi , \eta \to 0$.

Consider now the graphs  of figures 1a and 1b.
When we insert the spurion energies, the total energy $E_1+E_2$ on
the cut lines is not equal to the initial energy $E$, because there
is no delta function $\delta (E_1+E_2-E)$. But it is convenient
to insert the delta function $\delta (E_1+E_2+\Delta E-E)$
and remove it again by integrating over $\Delta E$.
Since the switching-off functions
constrain the spurion energies to be small, we may expand this delta
function in a Taylor series:
$$
\delta (E_1+E_2+\Delta E-E)=\sum _{\nu}  {1\over{\nu !}} (\Delta E)^{\nu}
\delta ^{[{\nu}]} (E_1+E_2-E)
$$
Then the expressions for figures 1a and 1b give contributions to
the decay rate that are of  the form
$$
{{\lambda ^2}\over{2E}}\int {{d^4p_1}\over{(2\pi )^4}} 2\pi\delta ^{(+)}
 (p_1^2-m^2)
{{d^4p_2}\over{(2\pi )^4}} 2\pi\delta ^{(+)} (p_2^2-m^2)
(2\pi )^3\delta ^{(3)}({\bf p}_1+{\bf p}_2-{\bf P})
\sum _{\nu} 2\pi \delta ^{[{\nu}]} (E_1+E_2-E)I^{\beta}_{\nu}(p_1)
$$ $$
\phantom{xxxxxxxxxxx}
={{\lambda ^2}\over{2E}}\int{{d^3p_1}\over{(2\pi )^3}}{1\over{4E_1E_2}}
\sum _{\nu} 2\pi \delta ^{[{\nu}]} (E_1+E_2-E)I^{\beta}_{\nu}(p_1)
\eqno(13)
$$
For figure 1a
$$
I^{\beta}_{\nu}(p_1)=N_1^{-1}\int {{d(\Delta E)}\over{2\pi}}
{(\Delta E)^{\nu}\over{\nu !}}
\int {{d\omega}\over{2\pi}}S_{\xi}(\Delta E-\omega )S^*_{\xi}(\Delta E)
S_{\eta}(\omega ){1\over{2E_1\omega+\omega ^2+i\epsilon}}
\Pi _a^{\beta}(p_1)
\eqno(14a)
$$
and for figure 1b
$$
I^{\beta}_{\nu}(p_1)= N_2^{-1}\int {{d\Delta E}\over{2\pi}}
{(\Delta E)^{\nu}\over{\nu !}}
\int {{d\omega}\over{2\pi}}\int {{d\omega '}\over{2\pi}}
S_{\xi}(\Delta E-\omega -\omega ')S^*_{\xi}(\Delta
E)\phantom{xxxxxxxxxxxxxxxxx}
$$ $$
\phantom{xxxxxxxxxxxxxxxxx}
S_{\eta}(\omega )S_{\eta}(\omega ')
{1\over{2E_1(\omega +\omega ')+(\omega +\omega ')^2+i\epsilon}}
\Pi _b^{\beta}(E_1+\omega ,{\bf p}_1)
\eqno(14b)
$$
We must add on the complex conjugate of each of these, corresponding to the
self-energy insertion being made instead in the complex conjugate
amplitude. With the explicit forms (2) the integrations in (14) are
easy to perform by taking residues at the poles of the integrand. Then,
when $\xi ,\eta \to 0$ we find that for both figure 1a and figure 1b
we may write the result as
$$
I^{\beta}_0(p_1)=-{{\Pi ^{\beta}(p_1)}\over{2E_1^2}}
 +{{\pd \Pi ^{\beta} (p_1)}\over{\pd E_1^2}}$$
$$I^{\beta}_1(p_1)={{ \Pi ^{\beta}(p_1)}\over{2E_1}}$$
$$I^{\beta}_{\nu}(p_1)=0 {\hbox{\ \  for }} {\nu}>1
\eqno(15)
$$
(though note that for figure 1a $\Pi ^{\beta}$ is independent of $p_1$ and so
the derivative term vanishes). In the Appendix, we show that the results
(15) are general in that they do not require the explicit form (2) for the
switching functions.

(13) and (15) correspond to self-energy insertions in the line $p_1$;
there are similar terms for insertions in the line $p_2$.

The derivative term in $I_0^{\beta}$ is naturally interpreted as
the finite-temperature renormalisation
of the squared coupling $\lambda ^2$ by the self-energy insertion in
figure 1b. At zero temperature the self-energy is a function of
$p_1^2$ and the coupling renormalisation involves its derivative,
which is naturally regarded as being with respect to $p_1^2$, but
one could equally well use $E_1^2$ because the result is the same. At
finite temperature the heat bath breaks the Lorentz invariance and the
variable of differentiation is necessarily $E_1^2$.
The graph of figure 1c also may be thought of as just further contributing to
the change of the coupling $\lambda$.

The other terms in (15) shift the mass $m$. To see this, note
that the zero-temperature decay rate $\Gamma ^0$ is
$$
\Gamma ^0={{\lambda ^2}\over{2E}}\int {{d^4p_1}\over{(2\pi )^4}}
{{d^4p_2}\over{(2\pi )^4}}
(2\pi )^4\delta ^{(4)}({ p}_1+{ p}_2-{ P})
2\pi\delta ^{(+)} (p_1^2-m^2)
2\pi\delta ^{(+)} (p_2^2-m^2)
$$ $$
\phantom{xxxxxxxxxxxxxxxxxxxxx}
={{\lambda ^2}\over{2E}}\int{{d^3p_1}\over{(2\pi )^3}}{1\over{4E_1E_2}}
 2\pi \delta  (E_1+E_2-E)
\eqno(16)
$$
Because
$$
{{\pd}\over{\pd m^2}}{1\over{E_1E_2}}=
  -{1\over{2E_1E_2}}\left ({1\over{E_1^2}}+{1\over{E_2^2}}\right )
$$ $$
{{\pd}\over{\pd m^2}} \delta (E_1+E_2-E)
=\delta' (E_1+E_2-E)\left ({1\over{2E_1}}+{1\over{2E_2}}\right )
$$
when we insert into (13) the nonderivative term of $I_0^{\beta}$ and
$I_1^{\beta}$ from (15), together with the corresponding contributions
from the self-energy insertion instead being made in $p_2$, we obtain
a correction to the decay rate equal to
$$
\Delta\Gamma ^0={{\lambda ^2}\over{2E}}\int {{d^4p_1}\over{(2\pi )^4}}
{{d^4p_2}\over{(2\pi )^4}}
(2\pi )^4\delta ^{(4)}({ p}_1+{ p}_2-{ P})
\phantom{xxxxxxxxxxxxxxxxxxxxxxxx}
$$$$
\phantom{xxxxxxxxxxxxxxx}
\left \{\left [\delta m^2(p_1){{\pd}\over{\pd m^2}}2\pi\delta ^{(+)}
(p_1^2-m^2)\right ] 2\pi\delta ^{(+)} (p_2^2-m^2)+
(p_1 \longleftrightarrow p_2)\right\}
\eqno(17a)
$$
where
$$
\delta m^2(p)=\left [\Pi _a^{\beta}(p)+\Pi _b^{\beta}(p)\right ]_
 {p^2=m^2}
\eqno(17b)
$$
Note that, while the contribution to $\delta m^2$ from $\Pi _a^{\beta}(p)$
depends only on $\beta$, that from  $\Pi _b^{\beta}(p)$ depends also
on $|{\bf p}|$. The effect of the heat bath on the decay rate depends
on how fast the decaying particle is moving.

We have shown, then, that the first three graphs of figure 1 may be
thought of as changing the mass of the decay-product particles
and the decay coupling. This has not needed any new assumptions about
how to renormalise at finite temperature, but is a consequence of
standard zero-temperature field theory combined with statistical
mechanics. In addition, there are the graphs of figures 1d and 1e,
which each represent the
absorption of a real photon from the heat bath and the stimulated
emission of a real photon into it.
\bigskip
{\bf 3 Scalar electrodynamics}

Consider now the proper scalar electrodynamics, with zero-mass photons.
It is convenient to use the gauge $A_0=0$. Then the thermal parts (7)
of the real and virtual photon propagators become
$$
\Delta ^{\beta} _{ij} (k)=D ^{\beta}_{ij}(k)=
  \left (\delta _{ij}-{{k_ik_j}\over{{\bf k}^2}}\right )
  2 \pi \delta (k^2)f^{\beta}(k_0)
$$
They are purely transverse: in the real-time formalism the
longitudinal-photon propagator does not have a thermal part\ref{4}.
The self-energies (9) become
$$
\Pi _a^{\beta}(p_1)=e^2\int {{d^4k}\over{(2\pi )^4}}D_{ii}^{\beta}(k)
=\sixth e^2 T^2$$
$$
\Pi _b^{\beta}(p_1)=e^2\int {{d^4k}\over{(2\pi )^4}}\;4p_{1i}
D_{ij}^{\beta}(k)p_{1j}\;
{1\over{(p_1+k)^2-m^2+i\epsilon}}
\eqno(18)$$
where we have used
$$
\int _0^{\infty}{{x\,dx}\over{e^x-1}}=\sixth\pi ^2
$$
and written $1/\beta =T$.
Again, $\Pi _a^{\beta}$ is independent of $p_1$.

The $A_0=0$ gauge is particularly convenient because with it the spurion
energies do not appear in the numerator factors of the Feynman integrands,
only in the denominators. So the analysis of figures 1a and 1b is
similar to before.
However, there are some differences.
First, when we set $p_1^2=m^2$  as required by the $\delta$-function in (13),
we obtain
$$
\Pi _b^{\beta}(p_1)=e^2\int {{d^4k}\over{(2\pi )^4}}\;4p_{1i}
D_{ij}^{\beta}(k)p_{1j}\;
\left\{\hbox{P}\;{1\over{2p_1.k}}-i\pi\delta (2p_1.k)\right\}
\eqno(19)
$$

The integrand
for Re $\Pi_b^{\beta}$ is odd in $k$ and so the integral vanishes.
$\Pi _b^{\beta}$ also has an imaginary part,
which arises because the Bose distribution
contains an infinite number of zero-energy photons; it is proportional to $T$.
(In the previous section, where we pretended that the photon had a nonzero
mass, this imaginary part was zero.)
However, the imaginary part cancels when we add to figure 1b its complex
conjugate. Thus
the whole of the additional mass shift (17b) caused by the heat bath
comes from the seagull self-energy insertion, figure 1a.
This is peculiar to  scalar electrodynamics in the $A_0=0$ gauge.
Gauge invariance ensures that we obtain the same total thermal mass shift  for
scalar electrodynamics in other gauges, but it generally
does not come only from figure 1a: in other gauges
$\Pi _b^{\beta}(p)$ does not vanish at $p^2=m^2$.
Also the spurion
energies may appear
in the numerators of the various integrals. The same applies to
spinor electrodynamics, even in $A_0=0$ gauge.

The other point is that
the derivative term in $I_0^{\beta}(p_1)$ in (15) is now infrared divergent
when we set $p_1^2=m^2$ as required by the $\delta$-function in (13).
This infrared divergence is cancelled by the graph of figure 1d.
In order to handle the infrared divergences, we add together
the graphs before we take the limits $\xi, \eta\to 0$.
There are also infrared divergences in figures
1c and 1e; again they cancel each other. In each of the two cancellations a
finite piece is left over. This finite piece is a combination of the change
in the effective value of the decay coupling constant $\lambda$
and the absorption and stimulated
emission of real photons.

Because the mass shift $\delta m^2$ of (17b) arises just from $\Pi _a^{\beta}$,
which according to (18) is independent of momentum, the correction (17a)
to the zero-temperature decay rate $\Gamma ^0$ is just
$\delta m^2 {{\pd \Gamma ^0}\over{\pd m^2}}$.
The zero-temperature decay rate is
$$
\Gamma  ^0={{\lambda ^2}\over{8\pi ME}}\sqrt{\quarter M^2-m^2}
\eqno(20)
$$
and so the mass shift changes this
by an amount
$$\Delta \Gamma _a = \sixth e^2 T^2 {{\pd \Gamma ^0}\over{\pd m^2}}
=-{{\lambda ^2e^2T^2}\over{96\pi ME}}{1\over{\sqrt{\quarter M^2-m^2}}}
\eqno(21)
$$
The change in the effective value of $\lambda$ changes the rate by
\def\p{{\bf p}}\def\k{{\bf k}}
$$
\Delta \Gamma _{bc} = -{{\lambda ^2e^2}\over{2E}}{1\over{(2\pi )^5}}
\int d^4k f^{\beta}(k_0)\delta (k^2)d^4p_1\delta ^{(+)}(p_1^2-m^2)
d^4p_2\delta ^{(+)}(p_2^2-m^2)\delta^{(4)}(p_1+p_2-P)$$ $$
\phantom{xxxxxxxxxxxxxxxxxxxx}
\left [{{\p _1^2-{{(\p _1.\k )^2}/{\k ^2}}}\over{(p_1.k)^2}}+
{{\p _1^2-{{(\p _2.\k )^2}/{\k ^2}}}\over{(p_1.k)^2}}-
2\;{{\p _1.\p _2-{{\p _1.\k\; \p _2.\k}/{\k ^2}}}\over{(p_1.k)(p_2.k)}}
\right ]
\eqno(22)
$$
where $f^{\beta}$ is the Bose distribution function (8). In the square
bracket, the first two terms correspond to the self energy insertion
of figure 1b and similar graphs, while the third corresponds to figure
1c. The way in which the contributions from these figures distribute
themselves is gauge-dependent and the separate terms break Lorentz symmetry
not only because of the presence of the Bose distribution associated with
the heat bath, but also because
we have used the $A_0=0$ gauge. However, if we use the $\delta$-functions in
(22) we find that we can reduce the square bracket to the Lorentz scalar
$$
 \left [-\left ({{p_1}\over{p_1.k}}-{{p_2}\over{p_2.k}}\right )^2\right ]
\eqno(23)
$$
so that in the sum of the three terms the Lorentz invariance is broken only
by the heat bath. It is straightforward to verify that the same answer is
obtained for $\Delta\Gamma _{bc}$ in Feynman gauge.

The contributions $\Delta \Gamma ^0_{de}$
from figures 1d and 1e, and similar figures, sum to
an integral exactly the same as (22), except that it has the opposite sign
and
$$
\delta^{(4)}(p_1+p_2-P)\longrightarrow\delta^{(4)}(p_1+p_2+k-P)
\eqno(24)
$$
It cancels the infrared divergence in (22).
In order to handle this new $\delta ^{(4)}$-function, it is convenient
to make changes in the integration variables $p_1$ and $p_2$ so as to
bring it back to its original form. We choose
$$p_1 \longrightarrow p_1-{{p_2.k}\over{P.k}}k$$ $$
p_2 \longrightarrow p_2-{{p_1.k}\over{P.k}}k
\eqno(25)
$$
These changes of variable have unit Jacobian and
leave the square bracket (23) unchanged,
but they result in
$$
\delta ^{(+)} (p_1^2-m^2)  \delta ^{(+)} (p_2^2-m^2) \longrightarrow
\delta ^{(+)} \left  (p_1^2-m^2-{{2p_1.k\; p_2.k}\over{P.k}}\right )
\delta ^{(+)} \left  (p_2^2-m^2-{{2p_1.k\; p_2.k}\over{P.k}}\right )
\eqno(26)
$$
That is, the difference between $\Gamma _{bc}$ and $\Gamma _{de}$, apart
from one of overall sign, is just in their $\delta$-functions (26). If we
want a low-temperature expansion  in the temperature $T$, we may
expand this difference as a Taylor series. The first surviving term gives
$$
\Delta\Gamma _{bc} + \Delta\Gamma _{de}=
{{\lambda ^2e^2}\over{E}}{1\over{(2\pi )^5}}
\left ({{\pd}\over{\pd m^2}}\right )^2
\phantom{xxxxxxxxxxxxxxxxxxxxxxxxxxxxxxxxxxxxxxxxxxxxxx}
$$$$
\int d^4k f^{\beta}(k_0)\delta (k^2)d^4p_1\delta ^{(+)}(p_1^2-m^2)
d^4p_2\delta ^{(+)}(p_2^2-m^2)\delta^{(4)}(p_1+p_2-P)
\left [M^2{{p_1.k\; p_2.k}\over{(P.k)^2}}-m^2\right ]
$$
$$
\phantom{xxxxxxxxxxxxxxxxxxxxxxxxxxxxxxxxxxxxxxxxxxxxxxxxxxxx}=
{{\lambda ^2e^2T^2}\over{96\pi ME}}{1\over{\sqrt{\quarter M^2-m^2}}}
\eqno(27)
$$
In this approximation, the decay rate in the rest frame of the initial
particle does not depend on how fast it is
moving relative to the heat bath. (That is not true of the contributions
from the higher-order terms in the expansion of the product of
$\delta$-functions in (26), which yield higher powers of $T$.)

The result (27) for $\Delta\Gamma _{bc} + \Delta\Gamma _{de}$
exactly cancels $\Delta \Gamma _a$ in (21). That is, the mass-shift
effect on the decay rate is cancelled, up to terms of order $T^2$, by
the effects of the change in the effective decay coupling and of
absorption and stimulated emission, whatever the speed of the initial
particle.

This result is applicable for low temperatures, $T\ll \sqrt{\quarter M^2-m^2}$.
If, instead, the temperature is high
(though still $e^2T\ll \sqrt{\quarter M^2-m^2}$
in order that the perturbation expansion may make sense), we find that
$\Delta\Gamma _{bc} + \Delta\Gamma _{de}$ is much smaller than
$\Delta \Gamma _a$. To see this, note that from (22) and (26) we have
$$
\Delta\Gamma _{bc} + \Delta\Gamma _{de}=\int {{k\,dk}\over{e^{\beta k}-1}}
        J(k,P)
\eqno(28)
$$
where $J$ is even in $k$, finite at $k=0$, and vanishes as
$1/k^2$ for
large k. When $\beta$ is small we may approximate the denominator in (28)
by $\beta k$, so that (28) becomes $T$ times a convergent integral.
On the other hand\footnote{$^*$}{If we had similarly expanded the denominator
in the integral that gives $\Delta\Gamma _a$, we should have obtained $T$
times a divergent integral.}, the result (21) for $\Delta\Gamma _a$ is exact
for
all $T$. That is, at high temperature the mass shift (which in
$A_0=0$ gauge arises just from figure 1a) is the dominant effect, while
at low temperature it is cancelled by the other effects.

So far we have supposed that the heat bath contains only photons. If the
charged particles also are in thermal equilibrium, there is already an
effect independent of the interaction with the photons: when, as we have
been supposing, the charged particles are bosons, there is stimulation
of their emission into the heat bath. If the initial particle $P$ is at rest
relative to the heat bath, this effect multiplies $\Gamma _0$ by
$[1+f^{\beta}(\half M)]^2$. (For fermions, there is a change of sign, so
that the heat bath tends to suppress the decay.) The corrections in order
$e^2$ are partly zero-temperature renormalisation to the mass and decay
coupling in this stimulated emission rate $\Gamma _0$. The analysis in
the Appendix of the effect of a self-energy insertion applies also when
it involves a thermal charged-particle
propagator (though now one must take account of the field doubling of
real-time finite-temperature field theory\ref{3}),
and so there will be additional
temperature-dependent corrections to the charged-particle mass and to the
decay coupling. At low temperature all these effects are exponentially
small, so that the $O(e^2 T^4)$ correction to $\Gamma _0$ from the photons
in the heat bath is the dominant one.
\bigskip
{\bf 4 Conclusions}

We have shown that, in order $e^2$,
the presence of a heat bath induces a mass shift,
proportional to the square of the temperature,
for charged particles that are involved in a reaction in the heat bath.
However, this mass shift cannot be dissociated from other effects,
including vertex corrections and stimulated
emission and absorption of real photons.
Gauge invariance results in cancellation of the infrared divergences that
are present in these separate effects. But at low temperature it also
leads to a cancellation of the mass-shift effect itself, so that the correction
to the reaction rate behaves only as the fourth power of the temperature.
On the other hand, at high temperature the mass shift is the dominant effect.

Our calculations are based on standard Feynman techniques of zero-temperature
field theory together with statistical mechanics. They are necessarily applied
to reaction rates rather than to amplitudes, and we find that field theory
at finite
temperature requires no new assumptions about renormalisation or asymptotic
conditions.
\bigskip
{\sl We are grateful to Antti Niemi for reading an early version of the
manuscript and to Tai Wu for reading a later one.
}

\bigskip
{\bf Appendix}

We derive the result (15) for the integrals (14) for general switching
functions $S_{\eta}(t)$ and $S_{\xi}(t)$, assuming only that they are real and
almost equal to unity for all finite $t$, but go to zero for infinite
$|t|$.

In (14a), which corresponds to figure 1a,
replace each of the switching functions $S$ by its Fourier-integral
representation and perform the $\Delta E$, $T'$ and $\omega$ integrations:
$$
I_{\nu}^{\beta}(p_1)={{-i}\over {2E_1}}
N_1^{-1}\int dT\,dtS_{\xi}(T)
\left [\left (-i{{\pd}\over{\pd T}}\right )^{\nu}S_{\xi}(T)\right ]S_{\eta}(t)
\left [\theta (t-T)e^{-2iE_1(t-T)}+\theta (T-t)\right ]\Pi ^{\beta}_a(p_1)
\eqno(A1)
$$
We must add  on the complex-conjugate integral.
Then, integrating by parts with respect to $t$ and using the definition
(12a) of $N_1$ we have
$$
I_0^{\beta}(p_1)=-{1\over{2E_1^2}}\Pi ^{\beta}_a(p_1)
\eqno(A2)
$$
together with a term proportional to
$$
\int dT\,dt [S_{\xi}(T)]^2{{\pd S_{\eta}(t)}\over{\pd t}}\cos (2E_1(t-T))
\eqno(A3)
$$
This integral is negligibly small: we may take the limit $\xi \to 0$
under the integral because the cosine ensures that the $T$ integration
converges
without this switching factor, and then vanishes. The $t$ integration vanishes
also.
The result (A2) agrees with (15), because for figure 1a $\Pi (p_1)$
is independent of $p_1$. By similar manipulations one finds that
$I_1^{\beta}(p_1)$ is equal to the result given in (15), together again with
a negligible term proportional to (A3), and that $I_{\nu}^{\beta}=0$ for
${\nu}>0$.

The corresponding analysis for the integral (14b), corresponding to figure 1b,
is only a little more complicated. Now $\Pi_b^{\beta}$ does depend on $p_1$.
Since for small $\eta,\xi$ only small spurion energies are important,
we may expand $\Pi_b^{\beta}(E_1+\o ,{\bf p_1})$ in a Taylor series in
powers of $\o$. With similar manipulations, we find that only the first
two terms in this expansion survive, and we obtain the results in (15).
Notice, however, that when the photon mass is taken to be zero, as it is
in section 3,  $\Pi_b^{\beta}(E_1+\o ,{\bf p_1})$ is not analytic at
$\o =0$ and so such a Taylor expansion is then suspect.

\bigskip
\bigskip
{\bf References}

1. J I Kapusta; {\it Finite-temperature field theory} (Cambridge University
Press, 1989); P V Landshoff and J I Kapusta, J Phys G15 (1989) 267

2. L Dolan and R Jackiw, Phys Rev D9 (1974) 3320;
Y Ueda, Phys Rev D9 (1981) 1383; G Peressutti and B S Skagerstam,
Phys Lett 110B (1982) 406; J L Cambier et al, Nuc Phys B209 (1982) 372;
M~Le~Bellac, 1991 Schladming lectures.

3. A Niemi and G W Semenoff, Nuc Phys B230 (1984) 181

4. K A James and P V Landshoff, Phys Lett B251 (1990) 167

5. A E I Johansson, G Peressutti and B S Skagerstam, Nuc Phys B278 (1986) 324

\bye

%There follows the main picture file.
%It inputs two other files, pictex.tex, which is a standard macro available
%on the bulletin board,  and pic.sav, which is attached (this is
%an intermediate file where some of the computations have been stored)
\voffset=2truein
\nopagenumbers
\input .pictex.tex
\setcoordinatesystem units <2truemm, 2truemm> point at 0 0
\beginpicture
\newbox\picA
\setbox\picA=\hbox{\beginpicture
\setcoordinatesystem point at 0 0
\setplotarea x from -17 to 17, y from -4 to 4
\linethickness=2pt
\replot "pic.sav"
\putrule from -17 0 to -9 0
\putrule from 17 0 to 9 0
%\ellipticalarc axes ratio 2:1 90 degrees from -9 0 center at -1 0
%\ellipticalarc axes ratio 2:1 -90 degrees from -9 0 center at -1 0
%\ellipticalarc axes ratio 2:1 90 degrees from 9 0 center at 1 0
%\ellipticalarc axes ratio 2:1 -90 degrees from 9 0 center at 1 0
\put {$p_1$} at  -2 5.5
\put {$p_2$} at -2 -5.5
\put {$P$} at  -15 1.5
\put {$>$} at  -1 4
\put {$>$} at  -1 -4
\put {$>$} at  -13 0
\endpicture}
\setcoordinatesystem units <2truemm, 2truemm> point at 50 -20
\put{\copy\picA} at 0 0
\put {(a)} at 0 -8.5
\put {$k$} at -5 1
\setdashes
\circulararc 360 degrees from -5 3.5 center at -5 1.7
\setsolid
\setcoordinatesystem units <2truemm, 2truemm> point at 10 -20
\put{\copy\picA} at 0 0
\put {(b)} at  0 -8.5
\put {$<$} at -6.5 0.45
\put {$k$} at -6.1 1.9
\setdashes
\circulararc -110 degrees from -2.5 3.7 center at -6.5 4.5
\setsolid
\setcoordinatesystem units <2truemm, 2truemm> point at 50 0
\put{\copy\picA} at 0 0
\put {(c)} at 0 -8.5
\put {$k$} at -4.5 0
\plot -6.5 .5 -6 -.2 -5.5 .5 /
\setdashes
\plot -6 3 -6 -3 /
\setsolid
\setcoordinatesystem units <2truemm, 2truemm> point at 10 0
\put{\copy\picA} at 0 0
\put {(d)} at 0 -8.5
\put {$k$} at -1.3 0
\put {$>$} at -1.3 1.75
\setdashes
\plot -6 3 -1 1.5 /
\plot 6 3 1 1.5 /
\setsolid
\setcoordinatesystem units <2truemm, 2truemm> point at 50 20
\put{\copy\picA} at 0 0
\put {(e)} at 0 -8.5
\put {$k$} at -2 -0.7
\plot -1.5 1.2 -1 0.5 -1.7 0.3 /
\setdashes
\plot -6 3 -1 .5 /
\plot 6 -3 1 -.5 /
\setsolid
\endpicture
\vskip 1truecm
\font\ten=cmr10 scaled\magstephalf
\centerline{{\ten Figure 1}}
\bigskip
\parindent=0pt
{\ten
Order $e^2$ diagrams for the decay $P \to p_1+p_2$ in a heat bath of
photons. The photon propagators $k$ correspond to the thermal function
(7). There are additional diagrams that are the complex-conjugate
of those shown, or have the r\^oles of $p_1$ and $p_2$ interchanged.
}
\bye